\documentclass[twocolumn,prl,superscriptaddress,amssymb,amsmath,floatfix,footinbib]{revtex4-2}
\usepackage{graphicx}
\usepackage{bm}
\usepackage{wrapfig}
\usepackage{times}
\usepackage[colorlinks=true,linkcolor=blue,urlcolor=blue,citecolor=blue]{hyperref}
\usepackage{float}

\begin{document}
\title{Tunable two-component ultracold molecular gases with vibrational shielding}

\author{Bijit Mukherjee}
\email{bijit.mukherjee@niser.ac.in}
\affiliation{School of Chemical Sciences, National Institute of Science Education and Research (NISER), Bhubaneswar, An OCC of Homi Bhabha National Institute, Khurda, Odisha 752050, India}
\author{Micha{\l} Tomza}
\email{michal.tomza@fuw.edu.pl}
\affiliation{Faculty of Physics, University of Warsaw, Pasteura 5, 02-093 Warsaw, Poland}

\date{\today}

\begin{abstract}
We propose a method to realize stable, tunable two-component quantum mixtures of ultracold polar molecules. First, we show that a pair of polar molecules in two distinct rovibrational states exhibits a repulsive interaction, thereby leading to collisional shielding without requiring any external field. We refer to this as ``vibrational shielding''. This intercomponent interaction is tunable by an external static electric or a microwave field, with the latter stabilizing both inter- and intracomponent interactions in a two-component bulk mixture. Additionally, we show that two microwave fields can independently tune the interactions of the individual components. Our findings thus open the door to the experimental realization of tunable quantum mixtures using polar molecules, analogous to tunable magnetic Feshbach resonances in two-component atomic quantum gases.
\end{abstract}

\maketitle

\textit{Introduction}--Quantum mixtures of ultracold atomic gases~\cite{Lamporesi2025} provide a versatile platform to study key physical phenomena ranging from miscibility of gases~\cite{TimmermansPRL98}, impurities in a quantum gas~\cite{SchirotzekPRL09}, to exotic topological structures~\cite{BaroniNatRevPhys24}. The relative intercomponent and intracomponent interactions in such systems are tuned via magnetic Feshbach resonances~\cite{ChinRMP2010}, enabling the realization of different configurations of the mixture. Ultracold polar molecules, on the other hand, have emerged as a powerful platform for studying many-body physics, quantum simulation, and quantum information~\cite{CarrNJP09, BaranovChemRev12, CornishNP24}. They offer stronger dipolar interactions~\cite{SchindewolfRMP26} than atoms and are expected to realize a wider variety of exotic quantum phases~\cite{MicheliNatPhys06, GorshkovPRL11, SchmidtPRR22, LangenPRL25, JinPRL25, CiardiPRL25}.

Recent breakthroughs in stabilizing ultracold molecular gases against collisional loss using \textit{shielding} techniques~\cite{AvdeenkovPRA06, WangNJP15, KarmanPRL18, LassablierePRL18, MukherjeePRR23, MatsudaScience20, AndereggScience21, KarmanPRXQ25} have enabled the creation of degenerate Fermi gases~\cite{SchindewolfNature22} and Bose-Einstein condensates (BECs) of dipolar molecules~\cite{BigagliNature24, ShiNatPhys26}. In addition, self-bound droplet phases of molecular BECs have been observed~\cite{ZhangNature26, ShiNatPhys26}. However, the realization of a tunable quantum mixture of molecular gases has remained elusive. Recent proposals on molecular SU($N$) magnetism~\cite{MukherjeeNJP25, MukherjeePRR25} and two-component fermionic gases~\cite{LiCommPhys26} predict collisionally stable mixtures, but no general method for independent tuning of the components exists.

\begin{figure}[tbh]
 \includegraphics[width=0.45\textwidth]{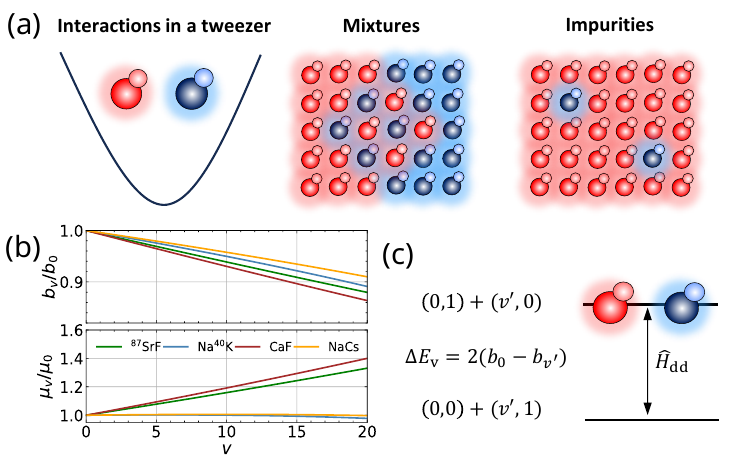}
\caption{(a) Cartoon showing three possible fields of study using two-component molecular mixtures. (b) Relative variation in $b$ and $\mu$ with vibrational level $v$ of a molecule. (c) Schematic showing the pair levels of interest $(v_1,n_1)+(v_2,n_2)$, their energy difference, and the coupling between them.}%
 \label{fig:schematic}
\end{figure}

In this letter, we theoretically propose an experimental scheme to realize highly tunable dipolar Bose-Bose and Fermi-Fermi quantum mixtures with ultracold molecular gases. First, we identify a pair of rovibrational levels of polar molecules as two internal states that exhibit a long-range, isotropic $R^{-6}$-type repulsive interaction, where $R$ is the intermolecular distance. This repulsive interaction leads to shielding against collisional loss without the need for an external field, similar to recent predictions that use rotational~\cite{WalravenPRA24}, hyperfine~\cite{WalravenPRA25}, and vibronic~\cite{MukherjeePRL26} states. We term this interaction as \textit{vibrational shielding}. Using this interaction with static electric or microwave fields, tunable quantum mixtures of ultracold molecular gases can be studied under three different perspectives: isolated two-body interactions in an optical tweezer, mixtures of two interacting gases, and single impurities immersed in a host quantum gas. This is depicted in a cartoon in Fig.\ \ref{fig:schematic}(a). Below, we describe the physics of vibrational shielding, methods for tuning inter- and intracomponent interactions while maintaining collisional shielding, and their potential applications.

\textit{Vibrational shielding}--We consider a pair of polar diatomic molecules that interacts in the long range through the dipole-dipole interaction given by $\hat{H}_\textrm{dd} = [\boldsymbol{\mu}_1 \cdot\boldsymbol{\mu}_2- 3(\boldsymbol{\mu}_1\cdot\hat{\boldsymbol{R}}) (\boldsymbol{\mu}_2\cdot\hat{\boldsymbol{R}})] / (4\pi\epsilon_0 R^3)$. Here $\boldsymbol{\mu}$ is the molecular permanent dipole moment, $\hat{\boldsymbol{R}}$ is a unit vector along the intermolecular axis, and $\epsilon_0$ is the vacuum permittivity. Identical molecules in rotational levels $n_1+n_2=0+1$ in the same vibronic state undergo resonant dipolar interaction due to a first-order coupling to the level $1+0$ via $\hat{H}_\textrm{dd}$. If the molecules, however, are prepared in different vibrational levels $v$, specifically in $(v_1,n_1)+(v_2,n_2)=(0,1)+(v',0)$, they undergo coupling via $\hat{H}_\textrm{dd}$ with the energetically different, but nearby level $(0,0)+(v',1)$. The energy difference originates from the fact that the molecular rotational constant $b$ (along with $\mu$) changes with $v$, and the energy difference between the two pair levels is $\Delta E_\textrm{v}=2(b_0-b_{v'})$. The subscripts denote vibrational levels, $v=0$ and $v'$.

The relative changes in $b$ and $\mu$ with $v$ for a few ultracold molecules of current experimental interest are shown in Fig.\ \ref{fig:schematic}(b). We use radial wavefunctions calculated from the ground electronic potentials~\cite{DocenkoEPJD04, Russier-Antoine:2000} for the molecules, and obtain the vibrationally averaged values $\mu_v$ and $b_v$~\footnote{The dipole moment functions for the alkali-metal diatoms are taken from Ref.\ \cite{LadjimiPRA24}. For CaF and SrF, we use the CCSD(T) method~\cite{BartlettRMP07} using MOLPRO~\cite{MOLPRO_brief} to calculate their ground electronic potentials and dipole moment functions with atomic basis sets and bond functions similar to Ref.\ \cite{LadjimiPRA24}.}. Usually, $b$ decreases with $v$, so that $\Delta E_\textrm{v}>0$, and molecules in the upper pair level $(0,1)+(v',0)$ experience an effective repulsion due to second-order dipolar coupling via the lower level $(0,0)+(v',1)$. This is shown schematically in Fig.\ \ref{fig:schematic}(c). Since $\hat{H}_\textrm{dd}$ scales as $R^{-3}$, the effective interaction scales as $C_6R^{-6}$, where $C_6= 2\mu^2_0\mu^2_{v'}/[9(4\pi\epsilon_0)^2\Delta E_\textrm{v}]$ (see End Matter). For the sake of universality, we introduce a reduced length scale $R_6=(1/2)(2\mu_\textrm{red}C_6/\hbar^2)^{1/4}$ and an energy scale $E_6=\hbar^2/(2\mu_\textrm{red}R_6^2)$, where $\mu_\textrm{red}$ is the reduced mass and $\hbar$ is the reduced Planck's constant. The energy gap $\Delta E_\textrm{v}$ is typically a few percent of $b_0$, if $v'=1-15$ (Table \ref{tab:properties} in End Matter). This gives rise to much larger values of $C_6$ compared to those originating from pure rotational states~\cite{WalravenPRA24}, and comparable to the ones due to hyperfine states~\cite{WalravenPRA25}. Note that we do not consider interactions arising from transition dipoles between inter-vibrational levels, as they are very small. 

\begin{figure}[tbh]
 \includegraphics[width=0.45\textwidth]{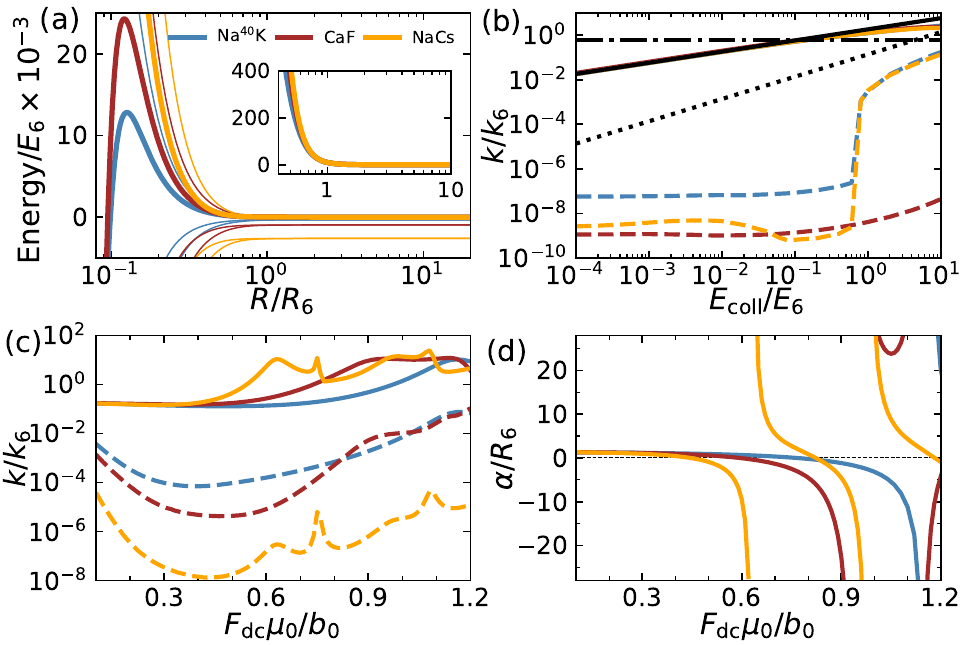}
\caption{(a) Adiabats for various molecules. The thick curves indicate the s-wave adiabat correlating to the upper pair level $(v,n)=(0,1)+(1,0)$. The thin curves indicate the remaining channels. The inset shows an expanded view of the long-range nature of the incoming adiabat. (b) Energy dependence of rates for intercomponent elastic scattering (colored solid lines) and loss processes (colored dashed lines). The black lines represent the universal rates for elastic scattering (solid) and intracomponent loss process (dash-dot for bosons and dotted for fermions) for a $C_6R^{-6}$ interaction. The colored solid lines are not visible as they lie below the black one. Panels (c) and (d) show rate coefficients and scattering lengths, respectively, as a function of static electric field at $E_\textrm{coll} = 0.01E_6$. }%
 \label{fig:adiabats_rate_coeff}
\end{figure}

We consider the adiabatic energies (hereafter ``adiabats'') involved in vibrational shielding. They are the $R$-dependent eigenvalues of the interaction Hamiltonian $\hat{H}_\textrm{int} = b_0\boldsymbol{\hat{n}}_0^2 + b_1\boldsymbol{\hat{n}}_1^2 +\hbar^2 \boldsymbol{\hat{L}}^2/(2\mu_\textrm{red}R^2) + \hat{H}_\textrm{dd}$, where the subscripts indicate two different vibrational levels, $\hat{\boldsymbol{n}}$ is the molecular rotation operator, and $\hat{\boldsymbol{L}}$ is the angular momentum operator for the relative rotation of the two molecules. Figure \ref{fig:adiabats_rate_coeff}(a) shows the adiabats for three molecules in reduced units. The curves are obtained for a fixed value of total angular momentum $J=1$ and parity $p=(-1)^{n_0+n_1+L}=-1$ such that the $L=0$ (s-wave) channel is included. The adiabat for the incoming s-wave channel for the pair level $(v,n)=(0,1)+(1,0)$ is repulsive at long range due to interactions with the lower-lying level $(0,0)+(1,1)$. It eventually becomes attractive at shorter range due to interactions with higher-lying pair levels, such as $(0,2)+(1,1)$. The resulting barrier heights are quite large compared to ultracold energies, leading to collisional shielding. The inset shows that at energies of up to a few hundred $E_6$, the interaction is universal for all molecules. For the sake of simplicity, we did not include the hyperfine channels in the above plot. However, our subsequent scattering calculations include them alongside $\hat{H}_\textrm{int}$ and the radial kinetic energy term.  

We carry out coupled-channel scattering calculations to evaluate rate coefficients for elastic scattering and loss processes. The details of the methodology are in the End Matter. Loss may occur from nonadiabatic transitions to the lower lying $(0,0)+(1,1)$ level, releasing kinetic energy $\Delta E_\textrm{v}$; or through a second-order transition via $(0,0)+(1,1)$ to a nearby energetically open hyperfine channel with much lower kinetic energy release. We consider three representative molecules in our calculations, namely, bosonic NaCs and CaF, and fermionic Na$^{40}$K, whose hyperfine energy splittings $\Delta E_\textrm{hf} \lesssim \Delta E_\textrm{v}$. Figure \ref{fig:adiabats_rate_coeff}(b) shows the calculated rate coefficients $k$ in terms of a reduced rate coefficient $k_6=4\pi R_6 \hbar/\mu_\textrm{red}$ (Table \ref{tab:properties}) as a function of reduced collision energies. The elastic rates are dominated by s-wave collisions that follow the universal threshold collision rate $k_{\textrm{el},L=0}=(32\pi^2k_6/\Gamma(\frac{1}{4})^4)(E_\textrm{coll}/E_6)^{1/2}$ for a $C_6R^{-6}$-type long-range potential~\cite{IdziaszekPRL10}. Here $E_\textrm{coll}$ is the collision energy. The interstate loss rates are suppressed by many orders of magnitude for $E_\textrm{coll}<E_6$. For the alkali-metal diatoms, the loss rates increase with higher energies $E_\textrm{coll} \gtrsim E_6$ due to spin-changing inelastic collisions. The above results are calculated for a magnetic field $B$ of 200 G, which decouples the hyperfine channels from the rotational manifolds. The initial hyperfine states for the molecules are chosen to be the lowest hyperfine-Zeeman component within the given rotational manifold. At low magnetic fields, the hyperfine-mediated loss increases, although not enough to destroy the shielding. The $B$-dependence of the loss rates is shown in Fig.\ \ref{fig:Bfield_rates} in the End Matter. Figure \ref{fig:adiabats_rate_coeff}(b) also shows the universal intracomponent loss rates~\cite{IdziaszekPRL10} for bosons and fermions. It is interesting to note that for fermions, the intercomponent elastic scattering is much faster than the intracomponent loss at low energies. From this, we can draw an important conclusion that vibrational shielding can effectively stabilize a 3D bulk mixture of fermionic molecules, thereby facilitating direct evaporative cooling. Our method thus provides a substantial improvement over the previous prediction of Ref.\ \cite{WalravenPRA24}, which relies upon forced evaporation in 2D confinements.

\textit{Effect of a static electric field}--Vibrational shielding leads to a purely repulsive interaction with a positive real part of the s-wave scattering length $\alpha$ of magnitude $\sim R_6$. An external static electric field $\boldsymbol{F}_\textrm{dc}$ can, however, impart an attractive dipolar interaction in the incoming pair level $(v,n,m_n)=(0,1,0)+(1,0,0)$, where $m_n$ is the projection of $n$ along $\boldsymbol{F}_\textrm{dc}$. Modest values of $F_\textrm{dc}$ allow the tuning of $\alpha$ across zero to large negative values, without destroying the collisional shielding. For highly dipolar molecules, like NaCs, $\alpha$ passes through poles. Figure \ref{fig:adiabats_rate_coeff}(c) shows the elastic and loss rate coefficients as a function of reduced electric fields. The values of $b_0/\mu_0$ are in Table \ref{tab:properties}, whereas the methods are in End Matter. We find that the rates of elastic scattering $k_\textrm{el}$ are enhanced by an order of magnitude due to enhanced dipolar interactions. This is evident from the poles in $\alpha$ [Fig.\ \ref{fig:adiabats_rate_coeff}(d)]. The loss rates $k_\textrm{loss}$, compared to the field-free case, increase due to field-induced nonadiabatic transitions to the pair levels $(0,1,\pm 1) + (1,0,0)$, even though they are much smaller than $k_\textrm{el}$. This method thus provides an improvement over the conventional static-field-shielding scenario, in which larger electric fields are required to achieve loss suppression~\cite{Gonzalez-MartinezPRA17, MukherjeePRR24}.

We envisage several applications of interactions arising from vibrational shielding in studies with optical tweezers. Firstly, both the zero-field interactions and their tunability with static fields can be used to implement near-deterministic loading of optical tweezers with single ultracold molecules, as proposed in Refs.\ \cite{WalravenPRL24, KarmanarXiv26}. Secondly, a pair of molecules can be loaded in a tweezer, exploiting the fact that intercomponent interactions form a dipolar blockade~\cite{WalravenPRL24} that protects the molecules against loss from the tweezer, whereas intrastate collisions lead to loss. These scenarios offer immense potential for studying quantum information processing with ultracold molecules~\cite{BaoScience23, HollandScience23}.

\textit{Effect of a microwave field}--In a 3D two-component mixture of quantum gases, the intra- and intercomponent collisions compete with each other. We have seen that for a fermionic mixture, vibrational shielding leads to much faster intercomponent elastic scattering than intracomponent loss processes at low energies. However, for bosons, which collide via the s-wave channel, the intracomponent loss rates are higher [Fig.\ \ref{fig:adiabats_rate_coeff}(b)]. In a single-component molecular 3D gas, microwave shielding has already proven instrumental in stabilizing molecules, particularly bosons, against collisional loss~\cite{BigagliNP23, LinPRX23}. Below, we show that a microwave field not only shields the molecules against intracomponent loss but also maintains intercomponent vibrational shielding. In addition, microwaves offer control knobs to tune interactions across all three pairwise components.

\begin{figure}[tbh]
 \includegraphics[width=0.45\textwidth]{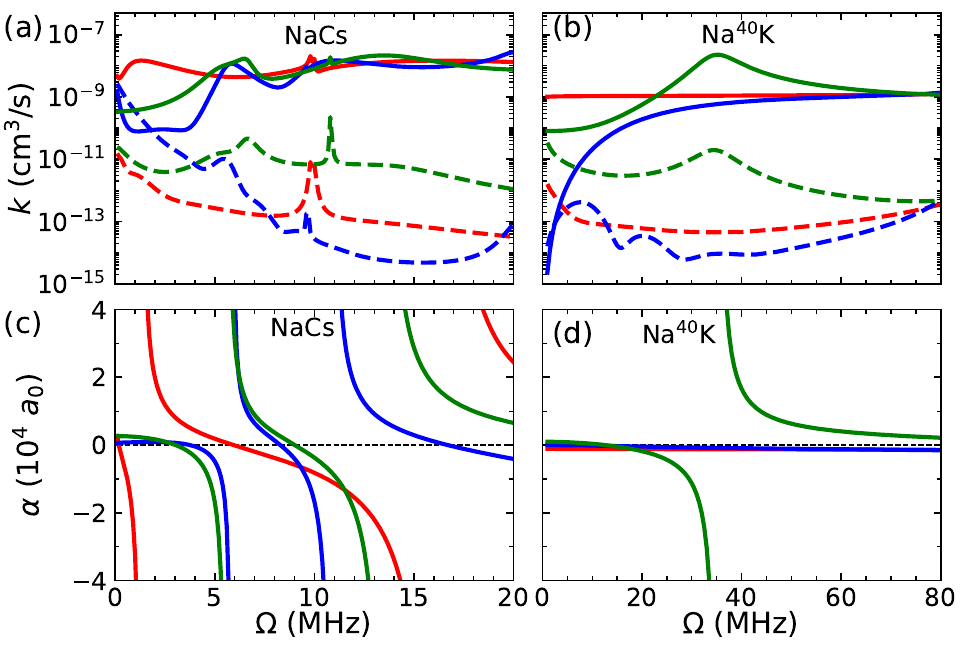}
\caption{Effect of a single $\sigma^+$-polarized microwave field in two-component mixtures of NaCs and Na$^{40}$K. The red (blue) curves correspond to the intracomponent $aa$ ($bb$) interactions, whereas the green curves denote the intercomponent $ab$ interactions. Panels (a) and (b) show rate coefficients for elastic scattering (solid lines) and loss processes (dashed lines), whereas panels (c) and (d) show scattering lengths for the corresponding components. For intracomponent scattering in Na$^{40}$K, we plot $\tilde{\alpha}_{11}$. The other quantities $\tilde{\alpha}_{10}$ and $\tilde{\alpha}_{1,-1}$ are similar in magnitude and therefore not shown.}%
 \label{fig:single_mw}
\end{figure}

We refer to the two component states $(v,n)=(0,1)$ and $(1,0)$ of the molecular mixture as $a$ and $b$, respectively. We apply a circularly polarized $\sigma^+$ microwave \textit{red-detuned} to the transition $n=0 \rightarrow 1$ for $a$ molecules. The microwave frequency is thus given by $\omega= 2b_0+\Delta$, where $\Delta$ is the microwave detuning. For shielding to occur for both intracomponent collisions $aa$ and $bb$, the detuning must satisfy 
\begin{align}\label{eq:cond_mw}
    \Delta<0 \quad \text{and} \quad |\Delta| \lesssim \Delta E_\textrm{v}.
\end{align}
This is because $a$ molecules are in the excited $n=1$ level, whereas $b$ molecules are in the $n=0$ level with a smaller rotational constant $b_1$. Collisions in $aa$ ($bb$) will experience antishielding if $\Delta>0$ ($|\Delta|>\Delta E_\textrm{v}$). The $b$ molecules experience a detuning $\omega-2b_1 = \Delta E_\textrm{v} + \Delta$, which is positive under condition (\ref{eq:cond_mw}). The two components also experience slightly different Rabi frequencies for the same microwave, namely, $\Omega$ for component $a$, and $\Omega \mu_1/\mu_0$ for component $b$.

We carry out coupled-channel scattering calculations in the presence of the microwave field for collisions in $aa$, $bb$, and $ab$ mixtures (methods in End Matter). Hereafter, all results are shown for a collision energy $E_\textrm{coll}/k_\textrm{B}=10$ nK ($k_\textrm{B}$ being Boltzmann's constant), which corresponds to a reasonably low temperature limit where quantum degeneracies are achieved. Figure \ref{fig:single_mw}(a) and (b) show $k_\textrm{el}$ and $k_\textrm{loss}$ for both intra- and intercomponent scattering in NaCs and Na$^{40}$K as a function of $\Omega$. We fix the microwave detuning-to-Rabi frequency ratio, $\delta=\Delta/\Omega$, to $-0.65$ for NaCs and $-0.15$ for Na$^{40}$K. These values satisfy condition (\ref{eq:cond_mw}) for the ranges of $\Omega$ values considered for each molecule. We find that for $\Omega \gtrsim 5$ MHz for NaCs and $\Omega \gtrsim 20$ MHz for Na$^{40}$K, the microwave stabilizes both intracomponent $aa$ and $bb$ collisions without destroying shielding in the intercomponent $ab$ collisions. Thus, a single microwave stabilizes a molecular mixture. The scattering properties are found to be weakly dependent on $\delta$ (Fig.\ \ref{fig:single_mw_EM} in End Matter).

The microwave field also allows tuning of the dipolar interactions among all three pairwise components. We study this by calculating the s-wave scattering lengths for collisions in all three pairwise components of NaCs and for the intercomponent collisions in Na$^{40}$K. For intracomponent collisions in Na$^{40}$K, we calculate the p-wave scattering length $\tilde{\alpha}_{LM_L}$~\cite{IdziaszekPRL10}, where $M_L$ is one of the projections of $L=1$ channel. Figure \ref{fig:single_mw}(c) and (d) show the plots of $\alpha$ as a function of $\Omega$. We find that for the Bose-Bose mixture of NaCs molecules, the scattering lengths $\alpha_{aa}$, $\alpha_{bb}$, and $\alpha_{ab}$ vary very differently. In contrast, for the Fermi mixture of Na$^{40}$K, $\alpha_{ab}$ is tunable across a pole. Leveraging tunability in $\alpha$ is possible by the application of two $\sigma^+$-polarized microwaves as discussed below.

It is worth noting that a single microwave results in a strongly interacting regime for a mixture of highly dipolar molecules such as NaCs. This is evident from the appearance of multiple poles in all $\alpha$ components. This may trigger inelastic recombination loss within each component. To avoid this, one may apply two microwave fields with elliptical and linear $\pi$ polarizations, thereby tuning the dipolar strength without forming two-body bound states~\cite{BigagliNature24, KarmanPRXQ25, ZhangNature26}. Alternatively, one may use a single microwave tuned to the transition $n=1 \rightarrow 2$ so that $\alpha$ is varied across zero without passing through a pole~\cite{DuttaarXiv26}.

\textit{Effect of two $\sigma^+$-microwaves}--A single microwave can address both components $a$ and $b$ if their $\Delta E_\textrm{v}$ is comparable to the microwave detuning $\Delta$. Thus, one has to limit the vibrational excitation to $v'=1$, as higher vibrations will destroy shielding in the $bb$ and $ab$ components. Also, a single microwave does not allow independent tuning of the two intracomponent interactions $aa$ and $bb$. Here, we propose that two $\sigma^+$-polarized microwaves simultaneously and independently tune both intracomponent $aa$ and $bb$ interactions. For this, we choose a highly excited vibrational state, say $v'=10$, as the second component, $b$. This ensures $\Delta E_\textrm{v} \sim 0.1b_0$ [see Fig.\ \ref{fig:schematic}(b)] to be much larger than individual microwave detunings. The two microwaves then independently address the molecular transition $n=0\rightarrow1$ for components $a$ and $b$, without interfering with each other.

\begin{figure}[tb]
 \includegraphics[width=0.45\textwidth]{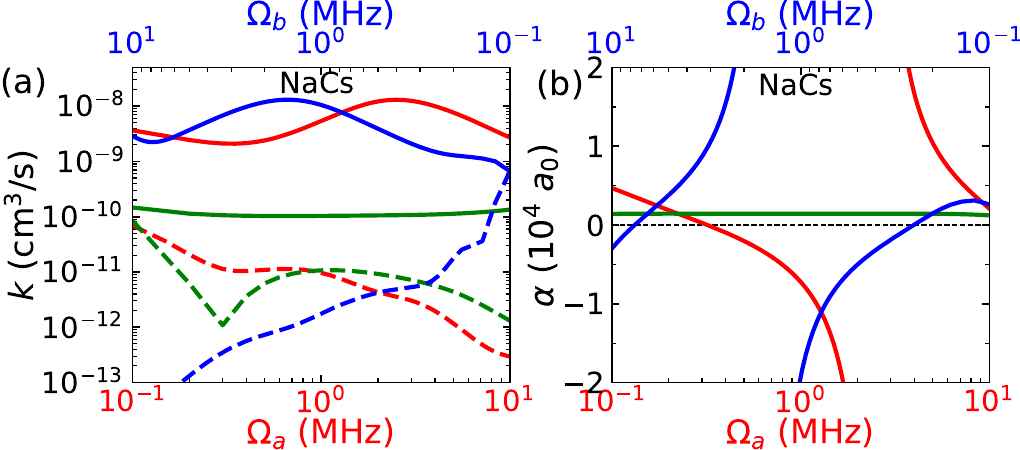}
\caption{Application of two $\sigma^+$-polarized microwave fields in an NaCs mixture. The red (blue) curves correspond to the intrastate $aa$ ($bb$) interactions, whereas the green curves denote the interstate $ab$ interactions. Panel (a) shows rate coefficients for elastic scattering (solid lines) and loss processes (dashed lines), whereas panel (b) shows scattering lengths for the corresponding components, for $E_\textrm{coll}/k_\textrm{B}=10$ nK. }%
 \label{fig:double_mw}
\end{figure}

Two microwave fields offer additional control knobs to tune the interactions in a molecular mixture, namely, $\Omega_a,\Delta_a$ and $\Omega_b,\Delta_b$. For intracomponent scattering, we consider the presence of both microwave fields, even though only one addresses a given component, with the other remaining far detuned. The intercomponent collisions, on the other hand, are affected by both fields. The details of the methods are in the End Matter. Figure \ref{fig:double_mw} shows the scattering properties of a shielded NaCs mixture in the presence of two microwaves. For simplicity, we have fixed three out of four parameters, namely, $\delta_a=\Delta_a/\Omega_a=-1$, $\delta_b=1$, and $\Omega_b=1/\Omega_a$, and varied $\Omega_a$. The intracomponent scattering properties follow similar trends as expected from a single microwave field and are universal across all polar molecules~\cite{DuttaPRR25}. The intercomponent interaction, however, is weaker than in the single-field case. This stems from the fact that the effective dipole moment $d_\textrm{eff} \propto 1/\sqrt{1+\delta^2}$ induced by the field that is far-detuned from the transition of one of the components is very small, and the dipolar interaction is proportional to the product of such effective dipoles. Nonetheless, shielding is maintained for intercomponent interactions, and the ratio $k_\textrm{el}/k_\textrm{loss}$ is expected to increase with temperature in the threshold regime~\cite{MukherjeePRR23}. Figure \ref{fig:double_mw_EM} in End Matter shows an overview of $k_\textrm{loss}$ in $ab$ collisions for various microwave parameters.

\textit{Study of quantum mixtures}--In an atomic Bose-Bose mixture, the sign and magnitude of the ratio $\gamma=\alpha_{ab}/\sqrt{\alpha_{aa}\alpha_{bb}}$ dictate the dynamics within a two-component quantum gas~\cite{BaroniNatRevPhys24}. For large and positive values of $\gamma$, repulsive interactions dominate, leading to immiscible mixtures~\cite{HallPRL81}. The components become miscible for $-1<\gamma<1$, whereas for small negative $\gamma<-1$, quantum droplets emerge~\cite{CabreraScience18}. Beyond this, a mean-field collapse occurs owing to dominating attractive intercomponent forces~\cite{PetrovPRL15}. Once $\gamma$ passes through a pole, few-body bound states form. In contrast, single-component shielded-molecular gases behave very differently. Strong dipolar coupling makes molecular BECs undergo a phase transition to dense self-bound droplets~\cite{LangenPRL25, JinPRL25, ZhangNature26, ShiNatPhys26} or droplet arrays~\cite{ZhangNature26}. Stronger Rabi couplings may induce a transition to a crystalline monolayer of molecules~\cite{CiardiPRL25}. Our results in Fig.\ \ref{fig:double_mw}(b) show that for a NaCs mixture, any value of $\gamma$ can be achieved by controlling one or the other microwave parameters. A two-component quantum mixture of such molecules is therefore expected to exhibit interesting yet distinct dynamics under our proposed scheme.

For Fermi mixtures, the physics is very different. Tuning $\gamma\sim k_\textrm{F}\alpha_{ab}$ ($k_\textrm{F}$ being the Fermi momentum) from small positive to negative values induces miscibility in the atomic mixtures~\cite{BaroniNatRevPhys24}. Beyond $\gamma<-1$, the components form Cooper pairs and then at the pole, undergo a transition from Bardeen–Cooper–Schrieffer (BCS) superfluid of Cooper pairs to BEC of $ab$ molecules~\cite{Zwerger2012bcs}. Our results in Fig.\ \ref{fig:single_mw}(d) show that a single microwave induces a pole in the intercomponent $\alpha_{ab}$, whereas the intracomponent interactions, dictated by $\tilde{\alpha}_{aa}$ and $\tilde{\alpha}_{bb}$, are weak. This novel intercomponent interaction can therefore be leveraged to study the BCS-BEC crossover in fermionic molecular superfluid mixtures. 

\textit{Outlook}--We proposed methods to realize stable and tunable interactions in a two-component quantum mixture. Two microwaves enable independent tuning of interactions within individual components of a bosonic mixture, whereas a single microwave allows tuning $\gamma$ for the fermionic ones. Another interesting application is to engineer independent on-site interactions in an optical lattice~\cite{PetrovPRA14, CardarelliPRA16, NixonQST24}. A stabilized quantum mixture loaded in an optical lattice may lead to interesting many-body physics if the pairwise on-site interactions $U_{aa}$, $U_{bb}$, and $U_{ab}$ are tuned independently, which is possible via our two-microwave tuning method.

Ultracold polar molecules are typically prepared in the ground vibronic state. Experiments employ either the Feshbach magnetoassociation technique followed by stimulated Raman Adiabatic Passage (STIRAP)~\cite{BergmannRMP98,ChinRMP2010}, or direct laser cooling~\cite{ShumanNature10}. Our proposal requires molecules to be in excited vibrational levels, which can be achieved experimentally via Raman transitions for $v=0 \rightarrow 1$ or via infrared laser pulses. One may design a STIRAP sequence to get molecules to desired $v'$ levels in the ground electronic state from a weakly-bound Feshbach state~\cite{PolovyPRA20}. On the other hand, a STIRAP transfer of population from $v=0$ to a highly excited $v'$ level may be designed. Excitations can be selectively performed in tweezers or in bulk gases, creating two-component molecular mixtures. Beyond studies of isolated interactions in optical tweezers or dynamics in bulk mixtures, our method can also serve as an interesting platform for studying polarons~\cite{SchmidtPRL15}. This may be achieved by introducing a second component as an impurity in a bath of single-component molecular gas [Fig.\ \ref{fig:schematic}(a)]. In conclusion, our current methods pave the way for studying a range of interesting phenomena using molecular mixtures.

\begin{acknowledgements}
\textit{Acknowledgments}--We thank Tapan Mishra for valuable discussions. BM is grateful to NISER Bhubaneswar for computational facilities. MT gratefully acknowledges the European Union (ERC, 101042989 -- QuantMol) for financial support and the Poland’s high-performance computing infrastructure PLGrid (HPC Center: ACK Cyfronet AGH) for providing computer facilities and support (computational grant no.~PLG/2025/018938). 
\end{acknowledgements}

\textit{Note added}—During the preparation of this manuscript, we became aware of parallel and independent works~\cite{Feng_arXiv26, jozwiak_arXiv26, WalravenarXiv26} that also consider using different vibrational levels for collisional shielding and its application to molecular quantum mixtures and loading of molecular arrays.

\bibliography{references}

\setcounter{equation}{0} 
\renewcommand{\theequation}{E\arabic{equation}}

\section{End matter}

\textit{Value of $C_6$ for vibrational shielding}--The effective interaction in vibrational shielding gives rise to a repulsive long-range $C_6R^{-6}$ potential. To derive the value of $C_6$, we consider a coupled basis set $|((n_\textrm{A}, n_\textrm{B})N_\textrm{AB}L)JM \rangle$ for molecules in vibrational states A and B. Here $n$ is the molecular rotation, $N_\textrm{AB}$ is the resultant of $n_\textrm{A}$ and $n_\textrm{B}$, $L$ is the end-over-end orbital angular momentum, $J$ is the total angular momentum and $M$ is its projection along the space-fixed $Z$ axis. The matrix element of $\hat{H}_\textrm{dd}$ in this basis is given by
\begin{align}
& \Big \langle \big ( (n_\textrm{A}, n_\textrm{B})N_\textrm{AB},L \big ) JM \Big | \hat{H}_\textrm{dd} \Big |\big ((n'_\textrm{A}, n'_\textrm{B})N'_\textrm{AB},L' \big ) JM \Big \rangle 
   \nonumber \\
&= -\sqrt{30} \frac{\mu_\textrm{A} \mu_\textrm{B}}{4\pi \epsilon_0 R^3} (-1)^{n_\textrm{A} + n_\textrm{B} + N'_\textrm{AB} -M}
\nonumber \\
& \times
([n_\textrm{A}] [n'_\textrm{A}] [n_\textrm{B}] [n'_\textrm{B}] [N_\textrm{AB}] [N'_\textrm{AB}] [L] [L'] [J])^{1/2}
\nonumber \\
& \times
\begin{pmatrix}
    J & 0 & J\\
    -M & 0 & M
\end{pmatrix}
\begin{pmatrix}
    L & 2 & L'\\
    0 & 0 & 0
\end{pmatrix}
\begin{pmatrix}
    n_\textrm{A} & 1 & n'_\textrm{A} \\
    0 & 0 & 0
\end{pmatrix}
\begin{pmatrix}
    n_\textrm{B} & 1 & n'_\textrm{B} \\
    0 & 0 & 0
\end{pmatrix}
\nonumber \\
& \times 
\begin{Bmatrix}
    N'_\textrm{AB} & L' & J\\
    L & N_\textrm{AB} & 2
\end{Bmatrix} 
    \begin{Bmatrix}
        n'_\textrm{A} & n'_\textrm{B} & N'_\textrm{AB}\\
        1 & 1 & 2 \\
        n_\textrm{A} & n_\textrm{B} & N_\textrm{AB}
    \end{Bmatrix},
    \label{eq:H_dd_matelem}
\end{align}
where the symbol $[X]=(2X+1)$. Since the molecules are assumed to be in the pair state $(n_\textrm{A},n_\textrm{B})=(1,0)$, which undergoes coupling with $(0,1)$, the value of $C_6$ is
\begin{align}
    C_6 &= \frac{\big(\langle ((1,0)1,0)1,0| \hat{H}_\textrm{dd} | ((0,1)1,2)1,0 \rangle\big)^2}{\Delta E_\textrm{v}} \nonumber \\
    &= \frac{2\mu^2_\textrm{A}\mu^2_\textrm{B}}{9(4\pi\epsilon_0)^2\Delta E_\textrm{v}}. 
\end{align}
The values of $C_6$ along a few other key parameters for a few ultracold molecules of current experimental interest are shown in Table \ref{tab:properties}.

\begin{table*}[tbp]
\caption{Key properties of pairs of molecules interacting in $(v_1,n_1)+(v_2,n_2)=(0,1)+(1,0)$. \label{tab:properties}} \centering
\begin{ruledtabular}
\begin{footnotesize}
\begin{tabular}{ccccccccc}
Molecules & $\mu_0$ (D) & $b_0$ (MHz) & $b_0/\mu_0$ (kV/cm) & $\Delta E_\textrm{v}$ (MHz) & $C_6$ (a.u.) & $R_6$ ($a_0$) & $E_6/k_\textrm{B}$ (K) & $k_6$ (cm$^3$s$^{-1}$) \\ \hline
Na$^{40}$K & 2.71 & 2820 & 1.33 & 27 & $7.0 \times 10^7$ & 843   & $3.87 \times10^{-6}$ & $1.1 \times 10^{-9}$ \\
$^{87}$SrF & 3.55 & 7440 & 4.17 & 92 & $6.2 \times 10^7$ &  929  & $1.90 \times10^{-6}$ & $7.4 \times 10^{-10}$ \\
NaCs  & 4.52 & 1740 & 0.76 & 14 & $1.1 \times 10^9$ & 2080 & $2.58 \times10^{-7}$ & $1.1 \times 10^{-9}$ \\
CaF  &  3.13 & 10240 & 6.90 & 150 & $2.4 \times 10^7$  & 634  & $7.32 \times10^{-6}$ & $9.1 \times 10^{-10}$ \\
\end{tabular}
\end{footnotesize}
\end{ruledtabular}
\end{table*}  

\textit{Coupled-channel method for vibrational shielding}--We consider polar molecules under the rigid-rotor approximation with vibrationally averaged $b$ values. For a pair of such colliding molecules in vibrational states A and B, the Hamiltonian reads
\begin{align}\label{eq:Hampair}
  \hat{H} &= \frac{\hbar^2}{2\mu_\textrm{red}} \Big(-R^{-1}\frac{d^2}{dR^2}R + \frac{\hat{\boldsymbol{L}}^2}{R^2} \Big) + b_\textrm{A}\hat{\boldsymbol{n}}_\textrm{A}^2 + b_\textrm{B}\hat{\boldsymbol{n}}_\textrm{B}^2 \nonumber \\
  &+ \hat{h}_\textrm{hf} + \hat{h}_\textrm{Zeeman} + \hat{H}_\textrm{dd},
\end{align}
where $\hat{h}_\textrm{hf}$ and $\hat{h}_\textrm{Zeeman}$ are the hyperfine and Zeeman Hamiltonians. They are defined elsewhere~\cite{MukherjeePRR23, MukherjeePRR25}. 

In the coupled-channel approach, the total wavefunction $\Psi$ is expanded $\Psi (R, \boldsymbol{\hat{R}}, \boldsymbol{\hat{r}}_\textrm{A}, \boldsymbol{\hat{r}}_\textrm{B}, \xi) = R^{-1} \sum_i \Phi_i (\boldsymbol{\hat{R}}, \boldsymbol{\hat{r}}_\textrm{A}, \boldsymbol{\hat{r}}_\textrm{B}, \xi) \psi_i (R)$, where $\psi_i(R)$ are the radial wavefunctions, $\boldsymbol{\hat{r}}_{\textrm{A(B)}}$ is a unit vector along the axis of molecule A(B), and $\xi$ denotes the spin coordinates. We use single-molecule basis functions $\phi \equiv (n,m_n,m_s,m_i)$ for CaF, and $(n,m_n,m_{i1},m_{i2})$ for the alkali-metal diatoms, where $m_s$ and $m_i$ are the projections of electron and nuclear spin angular momentum onto the $Z$ axis. The pair basis functions are therefore $\Phi_i=\phi_\textrm{A}(\boldsymbol{\hat{r}}_\textrm{A},\xi_\textrm{A}) \phi_\textrm{B} (\boldsymbol{\hat{r}}_\textrm{B},\xi_\textrm{B}) Y_{LM_L} (\boldsymbol{\hat{R}})$.

Calculations with hyperfine states are costly, so we restrict our pair rotational basis functions $(n_\textrm{A},m_{n,\textrm{A}})+(n_\textrm{B},m_{n,\textrm{B}})$ to the following: $(0,0)+(1,0)$, $(0,0)+(1,\pm 1)$, $(1,0)+(0,0)$ and $(1,\pm 1)+(0,0)$. The neglected rotational pair functions are energetically far off compared to $\Delta E_\textrm{v}$ and $\Delta E_\textrm{hf}$, and thus have a negligible contribution. We supplement these functions with all hyperfine channels along with $L$ up to 4. Since molecules are in non-identical states, both even and odd $L$ contribute. However, $\hat{H}_\textrm{dd}$ conserves their parity, which makes calculations cheaper. 

We select the following hyperfine channels for the molecules as the choice for the initial states: $(v,n,m_n,m_s,m_i)=(0,1,1,-1/2,1/2)+(1,0,0,-1/2,1/2)$ for CaF, $(v,n,m_n,m_{i,\textrm{Na}}, m_{i,\textrm{Cs}})=(0,1,1,3/2,7/2)+(1,0,0,3/2,7/2)$ for NaCs, and $(v,n,m_n,m_{i,\textrm{Na}}, m_{i,^{40}\textrm{K}})=(0,1,1,3/2,-4)+(1,0,0,3/2,-4)$ for Na$^{40}$K. This choice ensures that the hyperfine channels are lowest in energy within a given rotational manifold at high magnetic fields. The molecular hyperfine parameters are taken from Ref.\ \cite{Childs:CaF:1981, WillPRL16, AldegundePRA17}. The projection $M_\textrm{tot} = m_{n,\textrm{A}} + m_{n,\textrm{B}} + m_{g,\textrm{A}} + m_{g,\textrm{B}} + M_L$ of the grand total angular momentum is conserved. Here $m_g=m_s+m_i$ for CaF, and $m_g=m_{i,1} + m_{i,2}$ for the alkali-metal diatoms. We considered all values of $M_\textrm{tot}=M_F-2$ to $M_F+2$ in our calculations. Here $M_F=m_{n,\textrm{A}} + m_{n,\textrm{B}} + m_{g,\textrm{A}} + m_{g,\textrm{B}}$ of the intial states.

\begin{figure}[tbh]
 \includegraphics[width=0.45\textwidth]{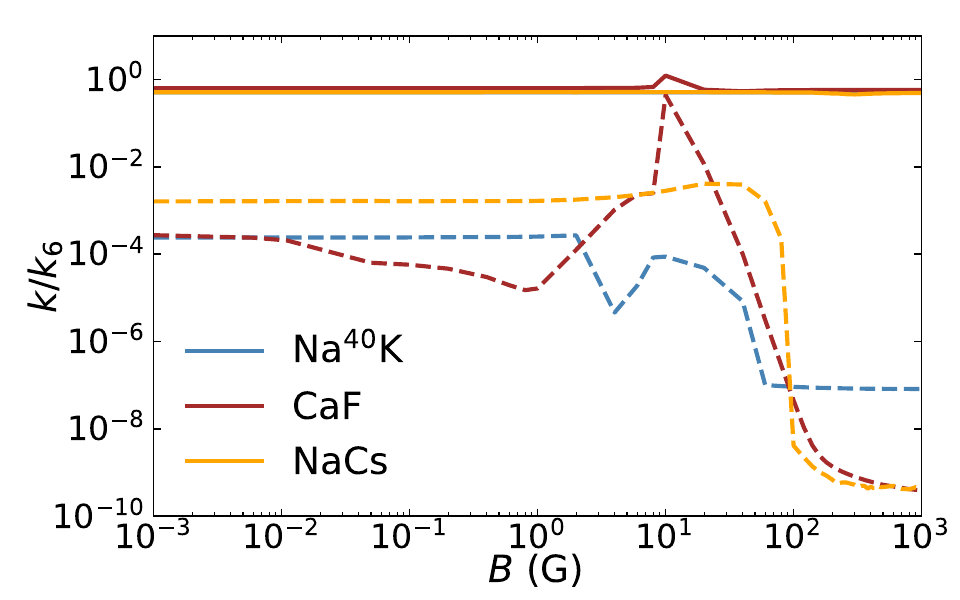}
\caption{Magnetic field dependence of the rate coefficients. Solid (dashed) lines indicate the elastic (total loss) rates. The collision energy is set to $0.1E_6$. }%
 \label{fig:Bfield_rates}
\end{figure}

We calculate cross sections for elastic scattering and loss processes as detailed in Ref.\ \cite{MukherjeePRR23}. We employ the MOLSCAT package~\cite{HutsonMolscatCPC19} for carrying out the coupled-channel scattering calculations using a fully absorbing boundary condition at the short range~\cite{ClaryFDCS87, JanssenPhD12}. We propagate the radial wavefunctions from $0.1R_6$ to $1000R_6$ in steps of $0.001R_6$. With the above basis set and propagation parameters, we obtain scattering results converged to within 1\%. The dependence of the rate coefficients on the magnetic field is shown in Fig.\ \ref{fig:Bfield_rates}.

\textit{Calculations with a static-electric field}--Modest electric fields are enough to decouple the rotational and hyperfine manifold without the need for magnetic fields. In such a case, the hyperfine states become spectators~\cite{MukherjeePRR25}, and we do not include them in our scattering calculations. We therefore adapt the spin-free coupled-channel approach of Ref.\ \cite{MukherjeePRR23}. We include field-dressed molecular rotor functions up to $\tilde{n}=3$ for each vibrational state A and B. The resulting pair basis set is divided into two groups: ``class 1'' and ``class 2''. The class 1 basis set includes pair-dressed functions: $(\tilde{n}_\textrm{A}, m_{n,\textrm{A}})+(\tilde{n}_\textrm{B}, m_{n,\textrm{B}})=(0,0)+(1,0)$, $(0,0)+(1,\pm 1)$, $(1,0)+(0,0)$ and $(1,\pm1)+(0,0)$. These are used explicitly in the coupled-channel calculations, while the class 2 functions are taken into account through Van Vleck transformations as described in Ref.\ \cite{MukherjeePRR23}. The initial pair state is chosen as $(1,0)+(0,0)$. Here, and in all our subsequent spin-free calculations, we include $L$ up to $12$, and $M_\textrm{tot}=-4$ to $4$.

\textit{Calculations with a single microwave field}--We follow the coupled-channel method of Ref.\ \cite{KarmanPRL18} for calculations of intracomponent $aa$ and $bb$ scattering. Microwave shielding requires a magnetic field $B\gtrsim100$ G to avoid hyperfine-mediated losses~\cite{KarmanPRL18}. A relatively high magnetic field makes the hyperfine states spectators, so we do not include them in our scattering calculations. For a $\sigma^+$-polarized microwave that dresses a pair of identical molecules A and B, we use the minimal rotor pair basis set: $(n_\textrm{A}, m_{n,\textrm{A}}, n_\textrm{B}, m_{n,\textrm{B}}, N)= (0,0,0,0,0)$, $(0,0,1,0,-1)$, $(0,0,1,\pm 1,-1)$, $(1,1,1,0,-2)$ and $(1,1,1,\pm1,-2)$, a total of 7 functions. Here $N$ denotes the photon number relative to a large bath of $N_0$ photons. We restrict the $L$-basis to even (odd) values for identical bosons (fermions).

For the calculation of intercomponent $ab$ collisions, molecules are in distinct quantum states. Hence, they require a larger rotor pair basis. We supplement the above list with the additional functions $(1,0,0,0,-1)$, $(1,\pm 1,0,0,-1)$, $(1,0,1,1,-2)$, and $(1,-1,1,1,-2)$. This brings the total rotor-pair functions to 12. In addition, we include both even and odd values of $L$ in our calculations.

\begin{figure}[tbh]
 \includegraphics[width=0.45\textwidth]{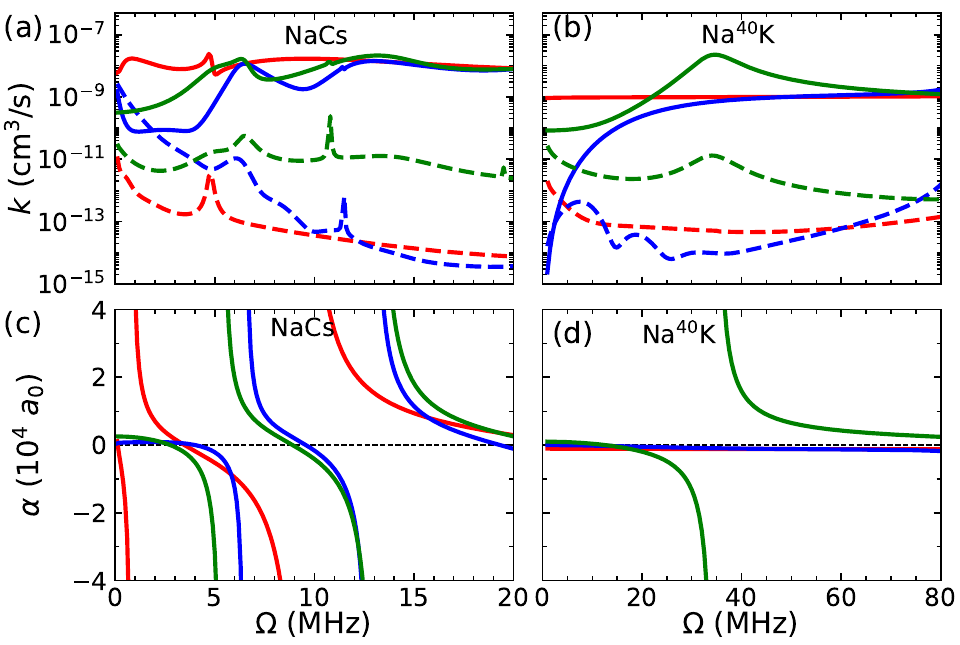}
\caption{Same as Fig.\ \ref{fig:single_mw}, except $\delta=-0.35$ for NaCs and $-0.27$ for Na$^{40}$K.}%
 \label{fig:single_mw_EM}
\end{figure}

Figure \ref{fig:single_mw_EM} shows the scattering properties of NaCs and Na$^{40}$K at different $\delta$ values from the ones chosen in the main text. This shows that the properties have a weak dependence on $\delta$ values permissible under condition (\ref{eq:cond_mw}).

\textit{Calculations with two microwave fields}--We follow the coupled-channel approach of Ref.\ \cite{KarmanPRXQ25} in which two microwave fields of different polarizations were employed. In our case, we have two fields of $\sigma^+$-polarization at different frequencies. The frequency difference $\Delta \omega$ between them is much larger than their individual detunings $\Delta$. This setup, therefore, avoids any ``Floquet'' driven inelastic loss due to photon-number-changing collisions~\cite{KarmanPRXQ25}.

Scattering with two microwaves requires larger basis sets. For calculations in intracomponent $aa$ and $bb$ collisions, we include rotor-pair functions $(n_\textrm{A}, m_{n,\textrm{A}}) + (n_\textrm{B}, m_{n,\textrm{B}}) + (N_\textrm{A}, N_\textrm{B})$: $(0,0)+(0,0)+(0,0)$, $(0,0)+(1,0)+(-1,0)$, $(0,0)+(1,\pm 1)+(-1,0)$, $(1,1)+(1,0)+(-2,0)$, $(1,1)+(1,\pm 1)+(-2,0)$, $(1,0)+(0,0)+(0,-1)$, $(1,\pm 1)+(0,0)+(0,-1)$, $(1,0)+(1,1)+(0,-2)$, $(1,\pm1)+(1,1)+(0,-2)$, giving rise to 13 functions. Here, $N_\textrm{A}$ and $N_\textrm{B}$ are photon numbers related to the two independent microwaves. We initiate the molecules in the level $(1,1)+(1,1)+(-2,0)$ for $aa$ collisions and $(0,0)+(0,0)+(0,0)$ for $bb$ collisions. Once again, here, we consider only even values of $L$ due to its conserved parity.

\begin{figure}[h]
\includegraphics[width=0.45\textwidth]{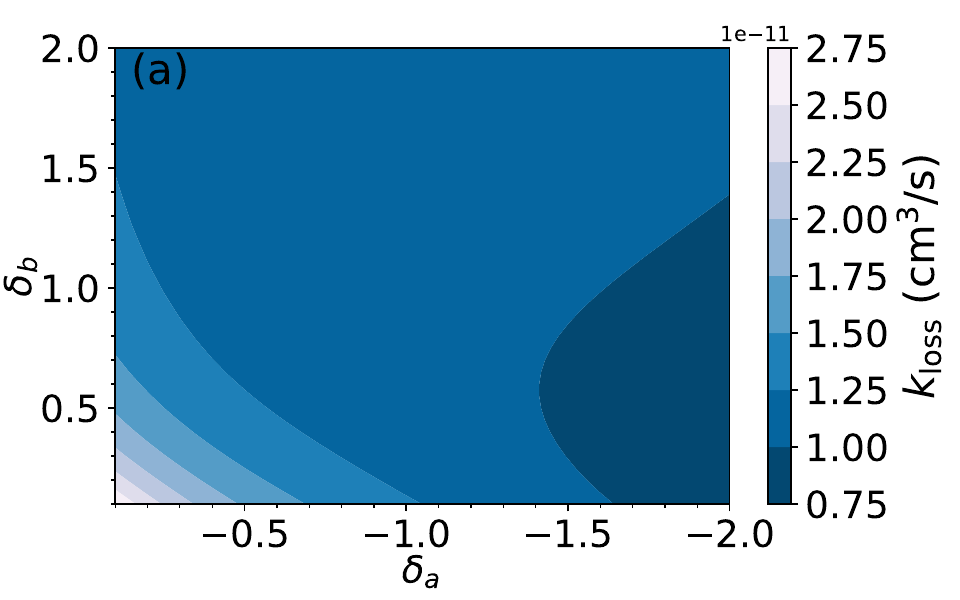}
\includegraphics[width=0.45\textwidth]{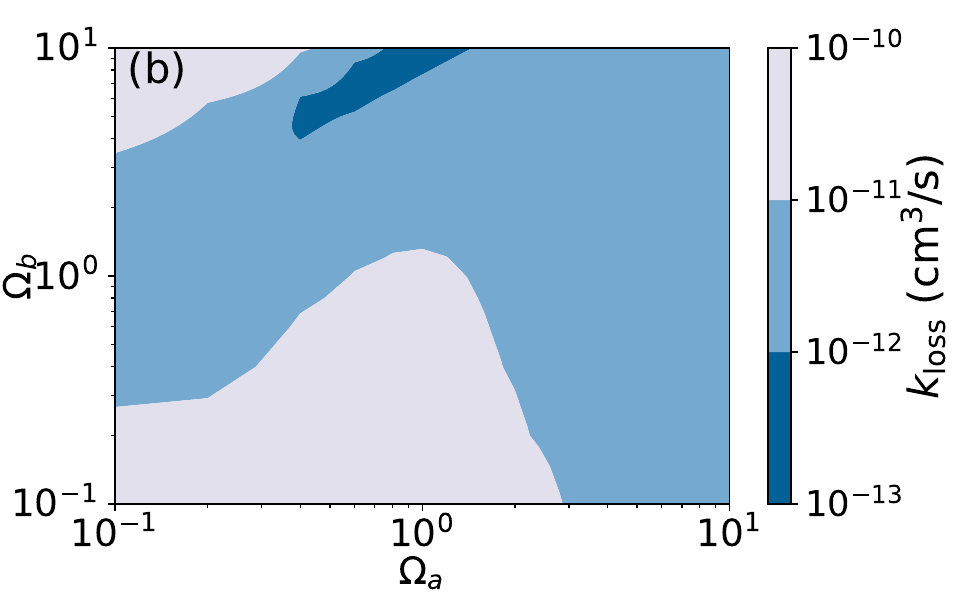}
\caption{Contour plots for loss rates $k_\textrm{loss}$ in intercomponent $ab$ collisions of NaCs mixture as functions of the two microwave parameters. The parameters fixed in panel (a): $\Omega_a=\Omega_b=1$ MHz, and in panel (b): $\delta_a=-\delta_b=-1$.}%
 \label{fig:double_mw_EM}
\end{figure}

For intercomponent scattering, we supplement the above list of functions with $(0,0)+(1,0)+(0,-1)$, $(0,0)+(1,\pm 1)+(0,-1)$, $(1,1)+(1,0)+(0,-2)$, $(1,1)+(1,-1)+(0,-2)$, $(1,0)+(0,0)+(-1,0)$, $(1,\pm 1)+(0,0)+(-1,0)$, $(1,0)+(1,1)+(-2,0)$, $(1,-1)+(1,1)+(-2,0)$, $(1,0)+(1,1)+(-1,-1)$, $(1,\pm1)+(1,1)+(-1,-1)$, $(1,1)+(1,0)+(-1,-1)$, and $(1,1)+(1,-1)+(-1,-1)$. This brings the total rotor-pair functions to 28. These levels are the closest in energy to the initial pair level $(1,1)+(0,0)+(-1,0)$. As usual, we include both even and odd values of $L$ in our calculations. Figure \ref{fig:double_mw_EM} shows the loss rate coefficient for the intercomponent scattering $ab$ in a NaCs mixture as a function of the two-microwave parameters.

\end{document}